\documentclass[pra,twocolumn,floatfix]{revtex4}
\usepackage{graphicx}
\usepackage{color}
\usepackage{amsmath, amsfonts, amssymb, bm}
\usepackage{braket}
\DeclareMathOperator*{\SumInt}{%
\mathchoice%
  {\ooalign{$\displaystyle\sum$\cr\hidewidth$\displaystyle\int$\hidewidth\cr}}
  {\ooalign{\raisebox{.14\height}{\scalebox{.7}{$\textstyle\sum$}}\cr\hidewidth$\textstyle\int$\hidewidth\cr}}
  {\ooalign{\raisebox{.2\height}{\scalebox{.6}{$\scriptstyle\sum$}}\cr$\scriptstyle\int$\cr}}
  {\ooalign{\raisebox{.2\height}{\scalebox{.6}{$\scriptstyle\sum$}}\cr$\scriptstyle\int$\cr}}}

\begin{document}
\title{Two-center electron-impact ionization via collisional excitation-autoionization}
\author{F. Gr\"ull}
\author{A. B. Voitkiv}
\author{C. M\"uller}
\affiliation{Institut f\"ur Theoretische Physik I, Heinrich Heine Universit\"at D\"usseldorf, Universit\"atsstr. 1, 40225 D\"usseldorf, Germany}
\date{\today}
\begin{abstract}
Electron-impact ionization of an atom or ion in the presence of a neighboring atom is studied. The latter is first collisionally excited by the incident electron, whose energy is assumed to be high but nonrelativistic. Afterwards, the excitation energy is transferred radiationlessly via a two-center Auger process to the other atom or ion, leading to its ionization. We show that the participation of the neighboring atom manifests in a very pronounced resonance peak in the energy-differential cross section and can substantially enhance the total cross section of electron-impact ionization. We also discuss the influence of the neighbouring atom on the angular distribution of the ejected electron. 
\end{abstract}
\maketitle

\section{Introduction}
Electron-impact ionization of atoms or ions is one of the most fundamental atomic collision processes \cite{book}. It is of importance, for instance, in various kinds of laboratory and astrophysical plasmas.
Apart from direct ionization by electron impact, there exist also more complex ionization mechanisms which involve autoionizing resonances. Upon the collision, a bound electron can be excited to an autoionizing state which subsequently stabilizes through Auger decay \cite{Peart,Hahn}. Besides, for certain resonant energies, a dielectronic capture of the projectile electron to the target may occur, leading to double Auger ionization \cite{Hahn,Muller}. These indirect ($e$,2$e$) processes, which rely on intra-atomic electron correlations, can be substantially more important than the direct ionization channel.

Autoionizing transitions can also arise from electrons situated at two different atomic centers, with Penning ionization being a famous example representing such an interatomic correlation effect.
In recent years, there have been extensive studies on another process driven by electron correlations in two or more atoms. An atom neighbouring another atom or ion in an excited state may receive the excitation energy radiationlessly via interatomic electron-electron interaction. Provided a sufficient energy transfer, the atom can then be ionized. Therefore, radiationless decay can happen in the system even when a single-center Auger decay is energetically forbidden. Despite its two-center nature, interatomic Coulombic decay (ICD)\cite{ICD, ICDres, ICDrev} as this process is called, can be much faster than the single-center radiative decay. Associated experiments have been carried out comprising noble gas dimers \cite{dimers}, clusters \cite{clusters} and water molecules \cite{water}. Both theoretical \cite{2CPI, Perina} and experimental \cite{2CPIexp,Hergenhahn} studies have been performed on the closely related process of two-center photoionization as well.

Photoabsorption serves as the common method to create the autoionizing state in experimental studies of ICD. Besides, a small number of experiments was carried out employing electron impact.  Using electron energies in the range of  3 keV \cite{Lanzhou}, 380 eV \cite{Lanzhou2} down to about 30--100\,eV \cite{Dorn}, ICD resulting from electron-impact ionization of one center, followed by molecular dissociation could be observed in noble-gas dimers and trimers. In these experiments there are, accordingly, three electrons in the final state: the primary scattered electron, the secondary ejected electron and the ICD electron. ICD was also implemented on water clusters adsorbed on condensed noble-gas surfaces using low-energy electron impact \cite{Grieves}.

\begin{figure}[b]  
\vspace{-0.25cm}
\begin{center}
\includegraphics[width=0.45\textwidth]{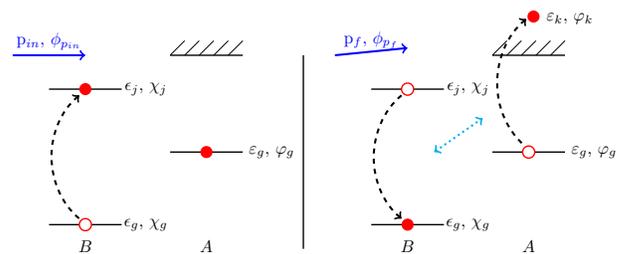}
\end{center}
\vspace{-0.5cm} 
\caption{Scheme of two-center electron-impact ionization 2C($e$,2$e$). A projectile electron first creates a two-center autoionizing state by collisionally exciting atom $B$ (left). Subsequently, the latter transfers the excess energy radiationlessly to atom $A$, leading to its ionization via two-center Auger decay (right).}
\label{figure1}
\end{figure}
A few electron impact induced interatomic processes have been treated theoretically so far. In two-center dielectronic recombination (2CDR) an incident electron is captured by an ion, leading to resonance excitation of a neighboring atom, which afterwards de-excites via spontaneous radiative decay \cite{2CDR}. If the metastable intermediate state does not stabilize radiatively, but instead re-emits the captured electron, two-center resonance scattering (2CRS) takes place \cite{2CRS}. In interatomic Coulombic electron capture (ICEC) the incident electron energy is so large that, upon its capture to the ion, a neighboring atom is ionized \cite{ICEC,q-dots}. The process thus represents an interatomic charge exchange. To our knowledge, comprehensive theoretical descriptions of interatomic {\it ionization} processes by electron impact, which take all steps of these processes into account, are still missing.  

In the present paper, we study electron-impact ionization via excitation-autoionization in a two-center atomic system consisting of atoms $A$ and $B$ -- a process which has not been considered in the literature yet. The autoionizing state of this system is formed by collisional excitation of atom $B$, which afterwards stabilzes via two-center Auger decay (or ICD), leading to ionization of a neighboring atom $A$ (see Fig.~\ref{figure1}). Note that, in contrast to Refs.~\cite{Lanzhou, Lanzhou2, Dorn}, the incident electron leads to excitation -- rather than ionization -- of atom $B$ and, consequently, there are only two electrons in the final state (scattered projectile and Auger electron).
We call this process resonant two-center electron-impact ionization 2C($e$,2$e$). It interferes with the usual direct electron-impact ionization of atom $A$. We will show that, due to 2C($e$,2$e$), electron-impact ionization can be qualitatively modified and strongly enhanced by several orders of magnitude in a narrow range of emitted electron energies. Resonant 2C($e$,2$e$) can also provide a substantial contribution to the total electron-impact ionization cross section of atom $A$.

Our paper is organized as follows. In Sec.~II we present our theoretical considerations of 2C($e$,2$e$). After formulating the general framework, an expression for the energy-differential cross section of this process will be obtained. Quantum interference effects from the different pathways will be included. In Sec.~III we illustrate our findings by some numerical examples and discuss their physical implications also including the angular distribution of 2C($e$,$2e$). As specific example, 2C($e$,2$e$) in a Li-He system at 1keV electron impact will be considered.
 Concluding remarks are given in Sec.~IV. Atomic units (a.u.) will be used throughout unless otherwise stated. The Bohr radius is denoted as $a_{0}$.

\section{Theoretical framework and analytical considerations}
\label{sec:theory}
\subsection{General treatment}
\begin{figure}[b]  
\begin{center}
\includegraphics[width=0.4\textwidth]{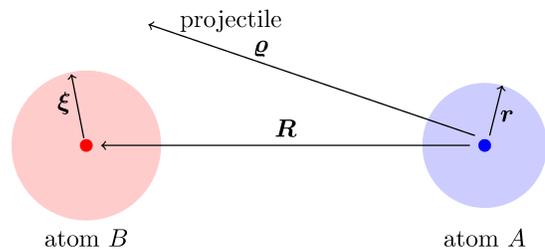}
\end{center}
\vspace{-0.5cm} 
\caption{Schematic representation of the spatial coordinates of the projectile electron and the atoms $A$ and $B$ with their corresponding active electrons.}
\label{Koordinaten}
\end{figure}

Focussing on the physical basics of 2C($e$,2$e$), we approach the process in a simple atomic system consisting of two atoms $A$ and $B$. These are initially in their ground states and separated by a distance $R$. In order to treat the two constituents individually and ignore molecular effects, $R$ has to be large enough, which also implies that the interaction between $A$ and $B$ is relatively weak. Within this consideration, 2C($e$,2$e$) can be viewed as a sequence of two sub-processes: The excitation of atom $B$ by electron-impact and the ionization of atom $A$ via dipole-dipole-interaction with atom $B$.
We suppose that both nuclei are at rest and carry charges $Z_{A}$ and $Z_{B}$ respectively.
In each atom only one electron is "active" in the considered process. We set the position of $Z_{A}$ as the origin and denote the coordinates of the  nucleus $Z_{B}$ (see Fig. \ref{Koordinaten}), the projectile electron, the electron associated with $A$ and the electron associated with $B$ by $\bf R$, ${\boldsymbol \varrho}$, $\bf r$ and ${\bf r'}={\bf R}+{\boldsymbol \xi}$, where $\boldsymbol \xi$ is the position of the electron of atom B with respect to the nucleus $Z_{B}$. We choose the $z$-axis to be along the incident electron momentum ${\bf p}_{in}=p_{in}{\bf e}_{z}$, also serving as our quantization axis.

Since the considered process involves two steps, we need to define the incident electron and initial, intermediate and final configurations of the two electrons associated with $A$ and $B$, which are illustrated in Fig. \ref{figure1}:\\
(I) $\Psi_{{\bf p}_{in},g,g}=\phi_{{\bf p}_{in}}({\boldsymbol \varrho})\varphi_{g}({\bf r})\chi_{g}({\boldsymbol \xi})$ with total energy $E_{{\bf p}_{in},g,g}=\frac{p_{in}^{2}}{2}+\varepsilon_{g}+\epsilon_{g}$. The initial state has an incident electron with momentum ${\bf p}_{in}$ and electrons of $A$ and $B$ in their ground states; (II) $\Psi_{{\bf p}'_{f},g,j}=\phi_{{\bf p}'_{f}}({\boldsymbol \varrho})\varphi_{g}({\bf r})\chi_{j}(\boldsymbol \xi)$
with total energy $E_{{\bf p}^{\prime}_{f},g,j}=\frac{p_{f}^{\prime 2}}{2}+\varepsilon_{g}+\epsilon_{j}$. In this intermediate state the incident electron has scattered and changed its momentum to ${\bf p}'_{f}$, the electron of $B$ has been excited and the electron in $A$ remains in its ground state; (III) $\Psi_{{\bf p}_{f},{\bf k},g}=\phi_{{\bf p}_{f}}({\boldsymbol \varrho})\varphi_{{\bf k}}({\bf r})\chi_{g}(\boldsymbol \xi)$
with total energy $E_{{\bf p}_{f},{\bf k},g}=\frac{p_{f}^{2}}{2}+\varepsilon_{k}+\epsilon_{g}$ with $\varepsilon_{k}=\frac{k^{2}}{2}$. The final state consists of the scattered electron, the electron from $A$, which  has been emitted into the continuum with asymptotic momentum $\bf k$ and the electron of $B$, which has returned to its ground state. Note that all continuum states are normalized to a quantization volue of unity.
In order to ionize atom $A$ in a two-center process including atom $B$, the ionization potential $I_{A}=|\varepsilon_{g}|$ has to be smaller than the energy difference $\omega_{B}=\epsilon_{j}-\epsilon_{g}$ of an electronic transition in atom $B$.\\
Assuming a sufficiently high projectile electron energy, the probability amplitude of 2C($e$,2$e$) is calculated via the second order of time-dependent perturbation theory,
\begin{align}
\label{eq: S2}
S^{(2)}=&-\int\limits_{-\infty}^{\infty}dt \displaystyle\SumInt  \mathcal{V}^{AB}({\bf p}'_{f},{\bf p}_{f},{\bf k})e^{-i(E_{{\bf p}'_{f},g,j}-E_{{\bf p}_{f},{\bf k},g})t}\nonumber\\
&\times\int\limits_{-\infty}^{t}dt' \mathcal{W}^{B}_{j}({\bf q})e^{-i(E_{{\bf p}_{in},g,g}-E_{{\bf p}'_{f},g,j})t'}.
\end{align}
The matrix elements are given by
\begin{eqnarray}
\mathcal{V}^{AB}({\bf p}^{\prime}_{f},{\bf p}_{f},{\bf k})&=&\langle \Psi_{{\bf p}_{f},{\bf k},g}|\hat{V}_{AB}|\Psi_{{\bf p}^{\prime}_{f},g,j}\rangle\\
\mathcal{W}^{B}(\bf q)&=&\langle\Psi_{{\bf p}^{\prime}_{f},g,j}|\hat{W}_{B}|\Psi_{{\bf p}_{in},g,g}\rangle ,
\end{eqnarray}
where ${\bf q}={\bf p}_{f}-{\bf p}_{in}$ is the momentum transfer experienced by the incident electron.\\
The interaction $ \hat{W}_{B}$ induces the electron-impact excitation of atom $B$ and is given by
\begin{eqnarray}
\label{eq:WB}
\hat{W}_{B}=-\frac{Z_{B}}{\left| {\boldsymbol \varrho}-{\bf R}\right|}+\frac{1}{\left|{\boldsymbol \varrho}-{\bf R}-{\boldsymbol\xi}\right|}
\end{eqnarray}
$\hat{V}_{AB}$ describes the two-center interaction between the electrons of $A$ and $B$, respectively
\begin{eqnarray}
\label{eq:VAB}
\hat{V}_{AB}=\frac{{\bf r}\cdot{\boldsymbol \xi}}{R^{3}}-\frac{3({\bf r}\cdot{\bf R})({\boldsymbol \xi}\cdot{\bf R})}{R^{5}}.
\end{eqnarray}
Here, we assume a dipole-allowed transition in atom $B$ and neglect retardation effects, which is justified for $R\ll c/\omega_{B}$.
We note besides that, as usual in scattering theory, the interactions 
(4) and (5) are assumed to be adiabatically switched on and off in the remote 
past and the distant future at $t\rightarrow \mp \infty$, respectively \cite{Wachter}.\\
Finally, $\displaystyle\SumInt=\int \frac{d^{3}p'_{f}}{(2\pi)^{3}}\sum_{j}$ is the integration over all intermediate continuum states of the scattered electron and the sum over all intermediate electron states of atom $B$.
Note that, since $\hat{V}_{AB}$ has no impact on ${\boldsymbol \varrho}$, one obtains ${\bf p}'_{f}={\bf p}_{f}$  due to the orthogonality of $\lbrace\phi_{{\bf p}'_{f}}({\boldsymbol \varrho}),\phi_{{\bf p}_{f}}({\boldsymbol \varrho})\rbrace$.\\
Integration  of Eq.~\eqref{eq: S2} over time yields
\begin{align}
\label{eq: S2-delta}
S^{(2)}&=-2\pi i\delta(E_{{\bf p}_{f},{\bf k},g}-E_{{\bf p}_{in},g,g})\frac{\mathcal{V}^{AB}({\bf p}_{f},{\bf k})\mathcal{W}^{B}({\bf q})}{\Delta+\frac{i}{2}\Gamma},
\end{align}
where $\Delta=\frac{p_{in}^2}{2}+\epsilon_{g}-\epsilon_{j}-\frac{p_{f}^{2}}{2}$ is the energy detuning and $\Gamma$ denotes the total decay width of the excited state $\chi_{e}({\boldsymbol \xi})$. It has been inserted to account for the finite lifetime of this state and includes the radiative width $\Gamma_{\text{rad}}$ as well as the two-center Auger width $\Gamma_{\text{aug}}$, often called ICD width, which are given by
  \begin{eqnarray}
  \label{eq: Gamma}
  \Gamma_{\text{rad}}^{B}&=&\frac{4 \omega_{B}^{3}}{3 c^{3}}\left|\left\langle\chi_{g}\left|{\boldsymbol \xi}\right|\chi_{j}\right\rangle\right|^{2}\\
  \Gamma_{\text{aug}}&=&\int \frac{d^{3}k}{(2\pi)^{2}}|\mathcal{V}^{AB}({\bf k})|^{2}\delta(\varepsilon_{k}+\epsilon_{g}-\varepsilon_{g}-\epsilon_{j}).
  \end{eqnarray}  
The $\delta$-function in Eq. \eqref{eq: S2-delta} expresses the energy conservation in the process:
\begin{eqnarray}
\label{eq: delta}
\left(\frac{p_{f}^{2}}{2}+\frac{k^{2}}{2}+\epsilon_{g}\right)-\left(\frac{p_{in}^{2}}{2}+\varepsilon_{g}+\epsilon_{g}\right)=0.
\end{eqnarray}
The ground state energy $\epsilon_{g}$ of atom $B$ effectively drops out from this expression since atom $B$ plays the role of a catalyzer in the process.

In order to obtain the differential cross section from Eq.~\eqref{eq: S2-delta}, we integrate the absolute square of the transition amplitude over the final momentum of the impacting electron ${\bf p}_{f}$ and the solid angle $\Omega_{k}$ of the emitted electron's momentum. Dividing this by the interaction time $\tau$ as well as the incident flux $j=p_{in}$, the differential cross section depends on the energies of the incident electron $\frac{{p}_{in}^{2}}{2}$ and ejected electron $\varepsilon_{k}$:
\begin{eqnarray}
\label{eq: sigma2}
\frac{d\sigma^{(2)}}{d\varepsilon_{k}}=\frac{1}{p_{in}}\int \frac{d^{3}p_{f}}{(2\pi)^{3}}\int \frac{k d\Omega_{k}}{(2\pi)^{3}}\frac{1}{\tau}\left|S^{(2)}\right|^{2}.
\end{eqnarray}
Our aim is to study the characteristics of this two-center process. Therefore, we compare it to the one-center process of electron-impact ionization of atom $A$.
Literature values for the one-center process of electron-impact ionization can easily be found for many atoms. Nevertheless, in order to be consistent with respect to the degree of approximation made, we calculate the cross section  of this process within the first order of time-dependent perturbation theory. The transition amplitude has the following form:
\begin{eqnarray}
\label{eq: S1}
S^{(1)}&=&-i\int\limits_{-\infty}^{\infty}dt\langle \phi_{{\boldsymbol p}_{f}}({\boldsymbol \varrho})\varphi_{{\bf k}}({\bf r})|\hat{W}_{A}| \phi_{{\boldsymbol p}_{in}}({\boldsymbol \varrho})\varphi_{g}({\bf r})\rangle\nonumber\\
& &\times e^{-i((\varepsilon_{g}+\frac{p_{in}^{2}}{2})-(\varepsilon_{k}+\frac{p_{f}^{2}}{2}))t}
\end{eqnarray}
with
\begin{eqnarray}
\hat{W}_{A}=-\frac{Z_{A}}{\left|\boldsymbol \varrho\right|}+\frac{1}{\left|{\boldsymbol \varrho}-\boldsymbol r\right|}
\end{eqnarray}
where ${\bf p}_{f}$, ${\bf p}_{in}$ and ${\bf k}$ are related due to energy conservation as shown in Eq. \eqref{eq: delta}.
This leads to the cross section
\begin{eqnarray}
\label{eq: sigma1}
\frac{d\sigma^{(1)}}{d\varepsilon_{k}}=\frac{1}{p_{in}}\int\frac{k d\Omega_{k}}{(2\pi)^{3}}\frac{1}{\tau}\left|S^{(1)}\right|^{2}.
\end{eqnarray}
The two-center channel competes with the direct electron-impact ioniziation. Furthermore, these two transitions interfere with each other and create a joint transition amplitude
\begin{eqnarray}
 S^{(1+2)}=S^{(1)}+S^{(2)}.
 \end{eqnarray} We refer to the resulting cross section  $\sigma^{(1+2)} $ as to the complete cross section for ionization of atom $A$ and obtain it analogously to Eq. \eqref{eq: sigma2}.
 Other competing channels may occur, such as impact-ionization of atom $B$. However, they do not interfere with the previoulsy established two, since they do not lead to the same final state. We will further refer to these channels in section \ref{Results}.
\subsection{Relation to single-center processes}
\label{single}
In this subsection we will restrict our consideration to ${\bf R}=R{\bf e}_{z}$ along the incident electron momentum ${\bf p}_{in}$. 
We note that taking an average over the orierentation of ${\bf R}$ does not change the features of 2C($e$,2$e$) qualitatively.

An advantage of the analytical approach to the two-center electron-impact ionization is the possibility of analyzing the mathematical form of the cross section. In order to enable comparison of our calculation to already known  quantities, we try to express $\frac{d\sigma^{(2)}}{d\varepsilon_{k}}$ as a composition of one-center processes.
Analysing the terms depicted in Eq. \eqref{eq: S2-delta} with the restriction to one intermediate state $j$, we can divide the expression into three one-center processes. The first step of electron-impact excitation of atom $B$ can be transferred into its according cross section $\sigma_{\text{exc}}^{B}$.  The dipole-dipole interaction can be separated into two processes involving atom $A$ and $B$ individually. The resulting matrix element concerning atom $B$ can be related to the radiative decay rate $\Gamma_{\text{rad}}$ as in Eq. \eqref{eq: Gamma} from the excited state back to the ground state $g$. For atom $A$ the matrix element can be expressed via the direct photoionization cross section $\sigma_{\text{PI}}^{A}$. 

A remark on the energy conservation conditions is appropriate. Within the composition consisting of one-center processes, we have two regulations of energy conservation: One in $\sigma_{\text{PI}}^{A}$ for $k$ and one in $\sigma_{\text{exc}}^{B}$ for $p_{f}$. In Eq. \eqref{eq: S2-delta} however,  we have only one $\delta$-function, in which both $k$ and $p_{f}$ are included.
In the present consideration we disentangle this combined law of energy conservation by assuming that, in the first step, $\frac{1}{2}\left(p_{f}^{2}-p_{in}^{2}\right)=\epsilon_{g}-\epsilon_{j}$ holds, whereas in the second step the electron in ejected from atom $A$ with energy $\frac{1}{2}k_{0}^{2}:=\varepsilon_{g}+\epsilon_{j}-\epsilon_{g}$. 

We then find the following compact formula for the differential cross section of 2C($e$,$2e$) at the resonance peak
\begin{eqnarray}
\label{eq: Vergleich}
\frac{d\sigma^{(2)}}{d\varepsilon_{k}}\Big|_{k=k_{0}}=\frac{2}{\pi}\frac{\Gamma_{\text{aug}}}{\Gamma^{2}}\sigma_{\text{exc}}^{B}(p_{in})
\end{eqnarray}
with the two-center Auger rate
\begin{equation}
\label{eq:Auger_Vergleich}
\Gamma_{\text{aug}}=\frac{3}{2\pi}\frac{c^{4}}{\omega^{4}R^{6}}\Gamma_{\text{rad}}^{(B)}\sigma_{\text{PI}}^{A}(\omega).
\end{equation}
Here, $\omega= \omega_{B}$ is the transition energy of atom $B$ in all quantities used in Eq. \eqref{eq: Vergleich} in the case of $k=k_{0}$. This expression enables a direct estimation of the cross section via one-center processes. For a fixed initial electron momentum $p_{in}$, the cross section decreases with $R^{-6}$ provided $R$ is large enough so that $\Gamma_{\text{rad}}\gg\Gamma_{\text{aug}}$ is valid. In this case, the dependence of $\Gamma$ on $R$ is negligible. For a fixed interatomic distance $R$, the cross section depends on the one-center processes of photoionization regarding atom $A$ and electron-impact excitation of atom $B$. As a result, 2C($e$,2$e$) will benefit if the cross sections of one-center photoionization of neighbour $A$ and one-center electron-impact excitation of neighbour $B$ are high.\\ We can now insert literature values for the expressions in Eq. \eqref{eq: Vergleich} and evaluate the ratio
\begin{eqnarray}
\label{eq:ratio}
\eta_{k_{0}}=\frac{d\sigma^{(2)}/d\varepsilon_{k}}{d\sigma^{(1)}/d\varepsilon_{k}}\Bigg|_{k=k_{0}}.
\end{eqnarray}
Let us consider a simple system consisting of H (as atom $A$) and He (as atom $B$). We choose an impact energy of  $\frac{p_{in}^{2}}{2}=250\text{ eV}$ and an internuclear separation of $R=20\,a_{0}$. At this distance and with $\sigma_{\text{PI}}^{\text{H}}=0.064\text{ a.u.}$ \cite{Samson}, one has $\Gamma_{\text{aug}}\approx\Gamma_{\text{rad}}=4.33\times 10^{-8}\text{ a.u.}$ \cite{NIST} for the dipole-allowed 1s-2p$_{0}$ transition in helium. Note that we have to multpliy the expression for $\Gamma_{\text{aug}}$ given in Eq. \eqref{eq:Auger_Vergleich} by a factor 2 in order to account for the two electrons in helium. This will be discussed more thoroughly in Sec. \ref{Results}. With $\sigma_{\text{exc}}^{\text{He}}(p_{in})=0.250\text{ a.u.}$ \cite{NIST} and $\frac{d\sigma_{\text{H}}^{(1)}}{d\varepsilon_{k}}|_{k=k_{0}}=1.94\text{ a.u.}$ \cite{Shyn}, we therefore obtain a ratio of $\eta_{k_{0}}\approx 4.7\times 10^{5}$. Thus the participation of the neighbouring helium atom tremendously amplifies the electron-impact ionization of hydrogen for ejected electron momenta close to $k_{0}$.
\\
When considering the total cross section however, one has to incorporate the behaviour for all $k$. The two-center process has a resonant nature in contrast to the one-center process.
 To highlight this difference we can approximate the two cross sections as follows:
\begin{eqnarray}
\label{eq: Vergleich_delta}
\sigma^{(2)}&\approx & \frac{d\sigma^{(2)}}{d\varepsilon_{k}}\Big|_{k=k_{0}}\delta_{\text{res}}\\
\sigma^{(1)}&\approx & \frac{d\sigma^{(1)}}{d\varepsilon_{k}}\Big|_{k=k_{0}}\delta_{\varepsilon_{k}}
\end{eqnarray}
Using the effective resonance width $\delta_{\text{res}}=\frac{\pi}{2}\Gamma$ which holds for a Lorentzian curve of the form $d\sigma^{(2)}/d\varepsilon_{k}\sim 1/\left(\Delta^{2}+\frac{1}{4}\Gamma^{2}\right)$, we obtain the total cross section of resonant 2C($e$,2$e$) in the form
\begin{equation}
\label{eq: sigma2_ana}
\sigma^{(2)}\approx\frac{\Gamma_{\text{aug}}}{\Gamma}\sigma_{\text{exc}}^{B}(p_{in}).
\end{equation}
This formula allows for a very intuitive interpretation. The first step in resonant 2C($e$,2$e$) is the electron-impact excitation of atom $B$, which enters into Eq.  \eqref{eq: sigma2_ana} by the corresponding cross section. The excited state may afterwards decay back into the ground state either by spontaneous radiative emission or by two-center Auger decay. The latter decay channel has a branching ratio of $\Gamma_{\text{aug}}/\Gamma$, which is reflected in Eq. \eqref{eq: sigma2_ana} as well.

In the H-He example under consideration, we find $\sigma^{(2)}_{\text{H-He}}\approx 0.12\text{ a.u.} $ for the resonant 2C($e$,2$e$) cross section.
 In comparison, the total cross section for one-center electron-impact ionization of H amounts to $\sigma^{(1)}=1.22\text{ a.u.}$ \cite{Lotz} at the same impact energy. Hence, the ratio of the fully-integrated cross sections is $\frac{\sigma^{(2)}_{\text{H-He}}}{\sigma^{(1)}_{\text{H}}}\approx 0.1$ at the chosen internuclear distance. The reason why the one-center process dominates over the two-center process in terms of the total cross section lies in their largely different effective widths. Using again $\frac{d\sigma_{\text{H}}^{(1)}}{d\varepsilon_{k}}|_{k=k_{0}}=1.94\text{ a.u.}$ \cite{Shyn} the value for the fully integrated cross section translates into an effective width $\delta_{\varepsilon_{k}}\approx 17\text{ eV}$  which is 7 orders of magnitude larger than $\delta_{\text{res}}$.

\begin{figure}\includegraphics[width=0.45\textwidth]{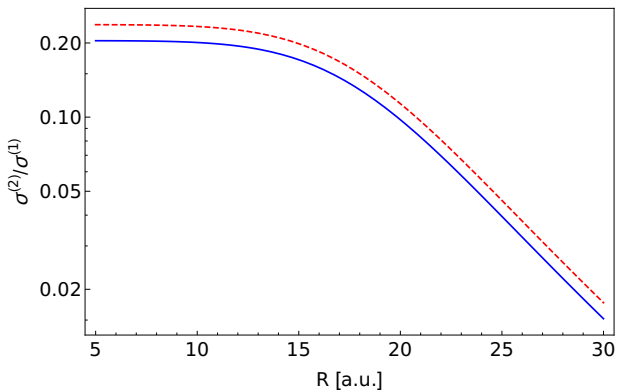}
\caption{Ratio $\frac{\sigma^{(2)}}{\sigma^{(1)}}$ calculated from literature values for H-He using $p_{in}^{2}/2=250\text{ eV}$ (blue solid) and $p_{in}^{2}/2=1\text{ keV}$ (red dashed). }
\label{fig:a}\end{figure}

Figure \ref{fig:a} shows the ratio $\frac{\sigma^{(2)}}{\sigma^{(1)}}$ of the total cross sections as function of the internuclear distance. The Lotz formula \cite{Lotz} has been used for $\sigma^{(1)}_{\text{H}}$, while Eq. \eqref{eq: sigma2_ana} has been evaluated to obtain $\sigma^{(2)}_{\text{H-He}}$. One can see that the cross section ratio saturates at distances $R\lesssim 12\,a_{0}$ where $\Gamma\approx \Gamma_{\text{aug}}$. In this region, the ratio attains a value of  $\sigma^{(2)}_{\text{H-He}}/\sigma^{(1)}_{\text{H}}\approx 0.2$, indicating a noteworthy relevance of the two-center process also for the total cross section of electron-impact ionization of hydrogen. We also show the ratio for $p_{in}^{2}/2=1\text{ keV}$ using $\sigma^{\text{He}}_{\text{exc}}\approx 0.11\text{a.u.}$ \cite{Ganas} since we will use this impact energy for our numerical calculations in Sec. \ref{Results}. The corresponding electron velocity is $v_{in}\approx 8.6 \text{ a.u.}$ where the applicability condition for perturbation theory ($v_{in}\gg 1\text{ a.u.}$) is well satisfied.

\section{Numerical results and Discussion}
\label{Results}
We want to further investigate the characteristics of resonant 2C($e$,$2e$) by illustrating our findings of the previous section by some numerical results. H-He as the simplest diatomic system was already considered in Sec. \ref{single}. However, H and He do not form a stable molecule. Therefore, we provide instead numerical results for a two-center system consisting of Li (as atom $A$) and He (as atom $B$). Both can form a $^{7}$Li$^{4}$He van-der-Waals molecule, which could be used for an experimental test of 2C($e$,$2e$) \cite{Friedrich}.

Since we consider one "active" electron for each atom, we choose effective nuclear charges $Z_{A}$ and $Z_{B}$ and electronic states $\chi$ and $\varphi$ in order to achieve a reasonable comparableness to real atomic species. Our goal is to ionize the electron associated with Li, so $\omega_{B}>I_{A}$ as described in Sec. \ref{sec:theory}. Using the single-active electron description, we describe lithium as a hydrogen-like atom with a 2s ground state.
 We choose the effective nuclear charge as $Z_{A}=1.259$ to match the binding energy $I_{A}\approx 5.39\text{ eV}$ of lithium \cite{NIST}. The neighbouring atom $B$ is chosen as helium in its ground state (1$\text{s}^{2}$). Within this example, we consider the dipole-allowed transition $(1s-2p_{m})$ for the electron-impact excitation. We use this transition in order to calculate a common effective nuclear charge for both states. With an excitation energy $\omega_{B}=\text{21.218 \text{ eV}}$, we find $Z_{B}=1.435$. 
 
We note that excitations to higher lying states (such as 3$p_{m}$, 4$p_{m}$ etc.) provide additional contributions to resonant 2C($e$,$2e$). However, we restrict our considerations here to the 2$p_{m}$ states in helium, since they have the largest dipole transition matrix elements from the ground state.
 
Our model now captures some basic features of a real Li-He dimer. However, we do not take account for any effects of the molecular bond.
As a van der Waals molecule, the dimer has a shallow interaction potential which allows for a large extension of the bond. Its mean distance $R \approx 55\,a_{0}$  exceeds the equilibrium distance $R_{\text{eq}}\approx 11.3\, a_{0}$, where the potential has a minimum \cite{Friedrich}.

The incident electron, before and after impact on $B$, is described by a plane wave $\phi_{{\bf p}_{in}}=e^{i{\bf p}_{in}{\boldsymbol \varrho}}$, $\phi_{{\bf p}_{f}}=e^{i{\bf p}_{f}{\boldsymbol \varrho}}$. After ionization, the influence of the remaining lithium ion on the emitted electron of $A$ is accounted for by using a Coulomb wave $\varphi_{{\bf k}}$ \cite{LandauQM} with the same nuclear charge $Z_{A}$ as mentioned above. We recall that all continuum states are normalized to a quantization volume of unity.

In Sec. \ref{sec:theory}, we established the theory for one "active" electron per atom. Considering helium as atom $B$, we want to describe it including both electrons. Therefore, instead of $\chi_{g}({\boldsymbol \xi})$ and $\chi_{j}({\boldsymbol \xi})$, we apply symmetrized wave functions:
\begin{align}
\label{eq: symmetrization}
\chi_{g}({\boldsymbol \xi}_{1},{\boldsymbol \xi}_{2})&=\alpha_{1s}({\boldsymbol \xi}_{1})\alpha_{1s}({\boldsymbol \xi}_{2})\\
\chi_{j}({\boldsymbol \xi}_{1},{\boldsymbol \xi}_{2})&=\frac{1}{{\sqrt{2}}}[\alpha_{j}({\boldsymbol \xi}_{1})\alpha_{1s}({\boldsymbol \xi}_{2})+\alpha_{1s}({\boldsymbol \xi}_{1})\alpha_{j}({\boldsymbol \xi}_{2})]
\end{align}
$\alpha_{1s}$ and $\alpha_{j}$ are hydrogenic wave functions with effective charge $Z_{B}$. Both interaction Hamiltonians in  Eq. \eqref{eq:WB} and \eqref{eq:VAB} are then properly extended according to
\begin{eqnarray}
\hat{W}_{B} &\rightarrow &- \frac{Z_{B}}{\left|{\boldsymbol \varrho}-{\boldsymbol R}\right|}+\sum_{\ell=1,2}\frac{1}{\left|{\boldsymbol \varrho}-{\boldsymbol R}-{\boldsymbol\xi}_{\ell}\right|} \\
\hat{V}_{AB} &\rightarrow &  \hat{V}_{AB}({\bf r},{\boldsymbol \xi}_{1}+{\boldsymbol \xi}_{2}).
\end{eqnarray}
These modifications lead to an additional factor of $2$ in the 2C($e$,$2e$) amplitude, as compared to the case when atom $B$ is an effective one-electron system. 

\subsection{Energy spectra}
\label{sec:Energy spectra}
First we consider 2C($e$,2$e$) as a separate channel and compare it to the one-center process.

As before, we restrict ourselves to ${\bf R}=R{\bf e}_{z}$. The internuclear distance is chosen as $R=20\,a_{0}$ and the incident electron energy as $\frac{p_{in}^{2}}{2}=1\text{ keV}$. At the chosen interatomic separation, the ratio $\Gamma_{\text{aug}}/\Gamma_{\text{rad}}\approx 0.2$ shows an excess of the radiative width compared to the Auger width. We point out that, for reasons of self-consistency, we use in our numerical calculations the value of $\Gamma_{\text{rad}}$ that follows from Eq. \eqref{eq: Gamma} when the wave functions in Eq. \eqref{eq: symmetrization} are inserted. Note in this regard that the matrix elements appearing in Eq. \eqref{eq: Gamma} for the decay width are also constituent parts of the 2C($e$,$2e$) cross section. Our calculated value of $\Gamma_{\text{rad}}$ is by a factor of $\approx 1.5$ larger than the literature value from \cite{NIST}. 

As we have already seen in Sec. \ref{single}, the 2C($e$,2$e$) singly differential cross section is very sensitive to ejected electron momentum $k$. For $k=k_{0}$ we obtain a peak displaying resonance \cite{Fussnote}. This represents a strong contrast to the one-center processes, where the lack of a resonance leads to smooth cross sections. 
 The values of the one-center cross section exceed those for the two-center process by several orders of magnitudes for most ejected electron energies $\varepsilon_{k}$. On resonance however, the two-center process increases tremendously and we obtain $d\sigma^{(2)}/d\varepsilon_{k}=2.4\times 10^{5}\text{ a.u.}$. For comparison, using instead Eq. \eqref{eq: Vergleich} and the literature values $\Gamma_{\text{rad}}\approx 4.33\times 10^{-8}\text{ a.u.}$ \cite{NIST}, $\sigma^{\text{Li}}_{\text{PI}}\approx 0.18\text{ a.u.}$ \cite{Bhatia} and $\sigma^{\text{He}}_{\text{exc}}\approx 0.11\text{ a.u.}$ \cite{Ganas} for $\frac{p_{in}^{2}}{2}=1\text{ keV}$, we find the approximate value  $\frac{d\sigma^{(2)}}{d\varepsilon_{k}}\Big|_{k=k_{0}}=2.5\times 10^{5}\text{ a.u.}$. 
 \begin{figure}\includegraphics[width=0.45\textwidth]{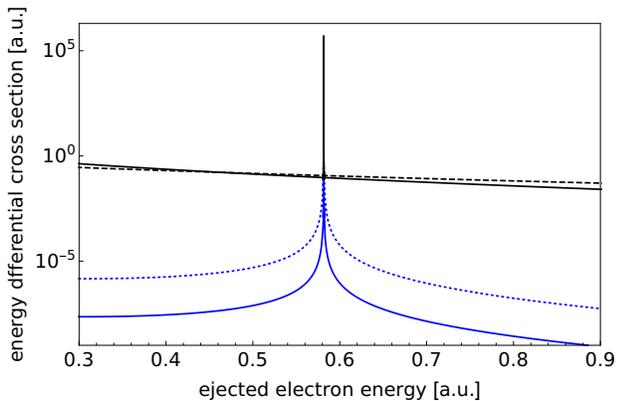}
\caption{Singly differential cross sections  of the one-center electron-impact ionization of lithium (black solid) $d\sigma^{(1)}/d\varepsilon_{k}$, two-center ionization of lithium $d\sigma^{(2)}/d\varepsilon_{k}$ for $R=20\,a_{0}$ (blue solid) and for $R=10\,a_{0}$ (blue dashed) with ${\bf R}\Vert {\bf p}_{in}$, and one-center ionization of helium (black dotted) $d\sigma^{(1)}/d\epsilon_{k}$  .}
\label{fig:Fig2}
\end{figure} 
Figure \ref{fig:Fig2} shows the correspondent singly differential cross sections $\frac{d\sigma^{(2)}}{d\varepsilon_{k}}$ and $\frac{d\sigma^{(1)}}{d\varepsilon_{k}}$ for a system representing Li-He at an internuclear distance $R=20\,a_{0}$.
\begin{figure}\includegraphics[width=0.45\textwidth]{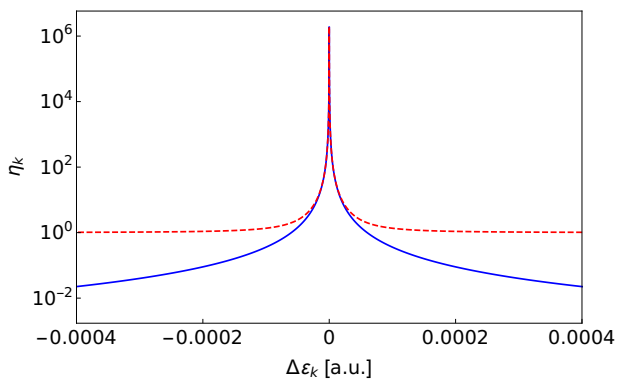}
\caption{Ratios $\eta^{(1+2)}_{k}$ (red dashed) and $\eta^{2}_{k}$ (blue solid) for $R=20\,a_{0}$ with $\Delta\varepsilon_{k}=k^{2}/2-k_{0}^{2}/2$.}
\label{fig:Versuch1}
\end{figure}  
Integrating in Fig \ref{fig:Fig2} over all possible $\varepsilon_{k}$ we obtain a value of  $\sigma^{(2)}_{\text{Li-He}}\approx  0.027\text{ a.u.}$.
For the one-center process our calculations using $Z=Z_{A}$ for the Coulomb wave function yield a total cross section of $\sigma^{(1)}_{\text{Li}}\approx 0.79\text{ a.u.}$ leading to a ratio $ \eta^{(2)}=\frac{\sigma^{(2)}_{\text{Li-He}}}{\sigma^{(1)}_{\text{Li}}} \approx 0.03$. We point out that our result for $\sigma^{(1)}_{\text{Li}}$ differs by a factor of $\approx 0.5$ from the Lotz value \cite{Lotz}. However,  since the same wave functions  for lithium are used in our 2C($e$,$2e$) calculations, we expect to slightly underestimate $\sigma_{\text{Li-He}}^{(2)}$ as well.

When the interatomic distance is reduced to $R=10\,a_{0}$, the ratio $\frac{\Gamma_{\text{aug}}}{\Gamma_{\text{rad}}}\approx 12.8$ shows the great significance of the two-center Auger decay for the total decay width, so $\Gamma\approx \Gamma_{\text{aug}}$. Inserting this approximation in Eq. \eqref{eq: sigma2_ana} yields $\sigma^{(2)}_{\text{Li-He}}\approx \sigma^{\text{He}}_{\text{exc}}\approx 0.11 \text{ a.u.}$ \cite{Ganas}. This way, we obtain $\eta^{(2)}\approx 0.07$ using $\sigma^{(1)}_{\text{Li}}$ from \cite{Lotz}. For the peak value our calculations yield $\frac{d\sigma^{(2)}}{d\varepsilon_{k}}\Big|_{k=k_{0}}=5.4\times 10^{5} \text{ a.u.}$. The corresponding value obtained from Eq. \eqref{eq: Vergleich} is smaller by a factor $\approx 0.12$.

We note that, in principle, the excited state in atom $B$ could be subject to fine structure splitting due to spin-orbit coupling. This effect would lead to an according splitting of the resonance peak, which is typically on the order of $10^{-4}\text{ eV}$ in light atoms. Since such a narrow doublet of lines is very difficult to resolve in electron-beam experiments, though, we refrain from its inclusion in our general treatment of resonant 2C($e$,$2e$). Besides, no fine structure arises in the excited $1s2p$ spin-singlet state \cite{NIST} in helium considered in the current section.
 
Figure \ref{fig:Fig2} also shows the cross section of the one-center electron-impact ionization of center $B$, in our case helium. As mentioned before, this process can compete but not interfere with the ionization of lithium. It was calculated by applying the same technique as above. The helium ground state was used in the form of  Eq. \eqref{eq: symmetrization} and the continuum state was described by the symmetrized wave function
\begin{equation}
\label{eq: Helium Couloumb}
\chi_{{\bf k}}({\boldsymbol \xi}_{1},{\boldsymbol \xi}_{2})=\frac{1}{\sqrt{2}}\left[\alpha_{{\bf k}}({\boldsymbol \xi}_{1})\alpha_{1s}({\boldsymbol \xi}_{2})+\alpha_{1s}({\boldsymbol \xi}_{1})\alpha_{{\bf k}}({\boldsymbol \xi}_{2})\right]
\end{equation}
where $\alpha_{\bf k}$ denotes a Coulomb wave \cite{LandauQM}. An effective nuclear charge $Z=1.34$ was applied which accounts for the correct first ionization energy of helium. This way we obtain results differing by a factor of $\approx 0.6$ from the values in \cite{Lotz} which fits in with the values for $\sigma^{(1)}$ seen above.
The fully integrated one-center cross section for lithium exceeds the one for helium by more than an order of magnitude \cite{Lotz}. The singly differential cross sections show this behaviour only for small $\varepsilon_{k}$. Around the resonance, one has $d\sigma^{(1)}_{\text{He}}/d\epsilon_{k}\approx d\sigma^{(1)}_{\text{Li}}/d\varepsilon_{k}$. We may conclude that the background of ejected electrons from impact ionization of helium as well as the total loss of neutral helium atoms by this process are not very severe.

\begin{figure}\includegraphics[width=0.45\textwidth]{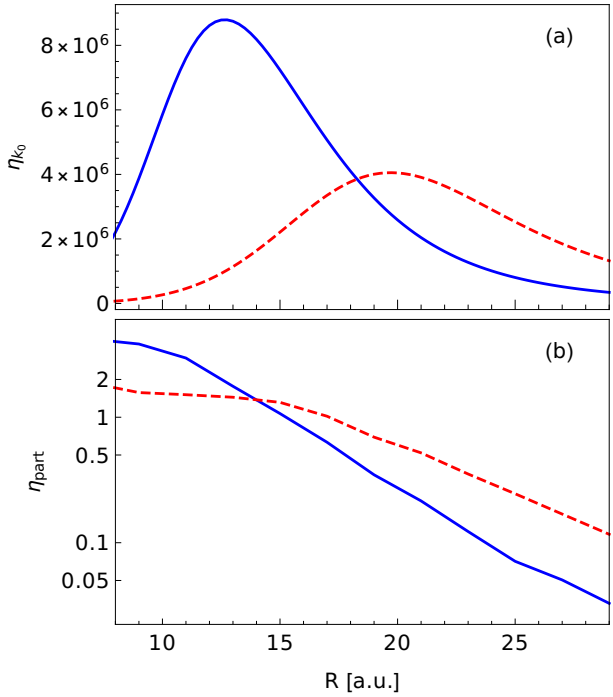}
\caption{(a) Ratios of the singly differential cross sections $\eta_{k_{0}}=\frac{d\sigma^{(2)}/d\varepsilon_{k}}{d\sigma^{(1)}/d\varepsilon_{k}}\Big|_{k=k_{0}}$ evaluated at the respective resonance for H-He (red dashed, calculated from literature values, multiplied by a factor of $20$) and Li-He (blue solid, calculated numerically).
(b) Ratios of the partial cross sections $\eta_{\text{part}}=\sigma^{(2)}_{\text{part}}/\sigma^{(1)}_{\text{part}}$ for H-He (red dashed) and Li-He (blue solid), taking an energy window of 1\,eV around the respective resonance into account.
 The incident energy is $\frac{1}{2}p_{in}^{2}=1\,\text{keV}$  in both panels.}
\label{fig:Figb}
\end{figure}

Next, we investigate the different characteristics of the two-center process and the complete two-center process involving quantum interference. Therefore, we plot the ratios $\eta^{(1+2)}_{k}=\frac{d\sigma^{(1+2)}/d\varepsilon_{k}}{d\sigma^{(1)}/d\varepsilon_{k}}$ and  $\eta^{(2)}_{k}=\frac{d\sigma^{(2)}/d\varepsilon_{k}}{d\sigma^{(1)}/d\varepsilon_{k}}$ calculated similarly to Eq. \eqref{eq:ratio}. 
In Fig. \ref{fig:Versuch1} the ratios of cross sections are depicted for a small range of energies  $\varepsilon_{k}=k^{2}/2$  around the resonance.
The ratio between the total cross section including the interference of both transition amplitudes $S^{(1)}$ and $S^{(2)}$ and the one-center process takes the constant value close to $1$ for ejected electron energies $\varepsilon_{k}$ off resonance. Therefore, the total cross section resembles the smooth curve of the one-center process for most values of $k$. For resonant $k$ the ratio experiences the peak which has already been observed in Fig. \ref{fig:Fig2}. The peak representating a tremendous amplification has a narrow width.
The calculation of the ratio comparing the two-center processes to the one-center process yields the peak behaviour on resonance which represents a strong amplification of the ionization for ejected electron momenta $k$ near resonance. 
We find a ratio $\eta_{k_{0}}=2.6\times 10^{6}$ at resonance when comparing the two-center process to the one-center process in Li-He at $R=20\,a_{0}$.
The $R$-dependence of this ratio is displayed in Fig. \ref{fig:Figb} (a). 

The resonance peaks shown in Figs. \ref{fig:Fig2} and \ref{fig:Versuch1} are so pronounced, that the underlying 2C($e$,$2e$) channel can give an appreciable contribution even to the total cross section of electron-impact ionization (see Fig. \ref{fig:a}). Because of their very narrow widths, however, the peaks themselves are very difficult to observe directly in experiment. Therefore, we have calculated the partial cross sections $\sigma^{(1,2)}_{\text{part}}$ which result from a resolvable energy range of ejected electrons around the resonance. For this purpose we have integrated the energy-differential cross sections over an interval of $1\text{ eV}$ width, including all electron energies $\varepsilon_{k}$ between $\frac{k_{0}^{2}}{2}\pm 0.5\text{ eV}$. Figure \ref{fig:Figb} (b) shows the dependency of $\sigma^{(2)}_{\text{part}}/\sigma^{(1)}_{\text{part}}$ on the internuclear distances $R$. Considering this restricted interval of energies, we obtain an even greater relevance of the two-center process than we have seen in Fig. \ref{fig:a}, reaching values $\eta_{\text{part}}> 1$. Consequently, the two-center process can dominate the electron-impact ionization in a measurable energy interval.

\subsection{Angular distributions}
We also want to investigate the angular distribution of the ejected electron in order to show differences between the two processes. 
To this end we calculate the doubly differential cross section $d^{2}\sigma^{(2)}/d\theta_{k}d\varepsilon_{k}$ at resonance, where the polar angle $\theta_{k}$ of the ejected electron is measured with respect to the momentum ${\bf p}_{in}=p_{in}{\bf e}_{z}$ of the incident electron. We choose $R=20\,a_{0}$ and consider two orientations, ${\bf R}_{\perp}=\frac{R}{\sqrt{2}}({\bf e}_{x}+{\bf e}_{y})$ and ${\bf R}_{\parallel}=R{\bf e}_{z}$, of the two-center system.
We point out that in a gas of Li-He dimers the internuclear separation vectors are randomly orientated. Hence, a prealignment of the molecular axes would be required to see the dependence on ${\bf R}$ of ejected electron's angular distribution in experiment.

\begin{figure}\includegraphics[width=0.45\textwidth]{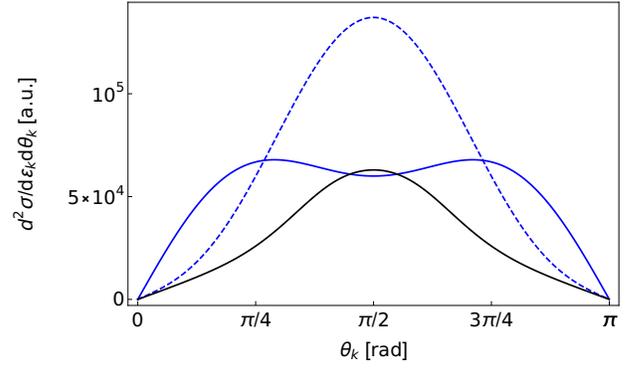}
\caption{Doubly differential cross sections $d^{2}\sigma/d\theta_{k} d\varepsilon_{k}|_{k=k_{0}}$ for one-center electron-impact ionization of lithium (black,  multiplied by a factor of $10^{6}$), and two-center ionization of lithium in a Li-He system, with interatomic orientations ${\bf R}_{\perp}=\frac{R}{\sqrt{2}}({\bf e}_{x}+{\bf e}_{y})$ (blue dashed) and ${\bf R}_{\parallel}=R{\bf e}_{z}$ (blue solid), respectively. The interatomic distance is $R=20\,a_{0}$, the incident energy $\frac{1}{2}p_{in}^{2}=1\text{ keV}$, and the emission energy $\frac{1}{2}k_{0}^{2}\approx 15.8\text{ eV}$.}
\label{fig:Fig4}
\end{figure}
In Fig. \ref{fig:Fig4} the angular dependencies of the cross sections are depicted. All curves are smooth and symmetric with respect to reflection at $\theta_{k}=\pi/2$. The angular distribution of the one-center process passes through a maximum at this angle. Because the plot shows the differential cross sections on resonance, the value of the direct pathway can by no means come up to the value of the two-center process. The latter depends on the orientation of ${\bf R}$. For ${\bf R}_{\perp}$, we obtain the position of the maximum at $\theta_{k}=\pi/2$ as well. While the overall shape looks similar, the slope on both sides slighty differs from the one-center case. Assuming instead an interatomic orientation along ${\bf R}_{\parallel}$ leads to a more drastic change of form. We then obtain a kind of plateau exhibiting two maxima and a shallow minimum at $\theta_{k}=\pi/2$. 

The two-center electron impact ionization thus noticably modifies the angular dependency of the ejected electron in contrast to the direct pathway. The  indirect process includes contributions from various excited electronic states in atom $B$. We further analyse their individual angular dependencies by calculating the contributions to the cross section stemming from each of the excited states separately.

\begin{figure}
\includegraphics[width=0.45\textwidth]{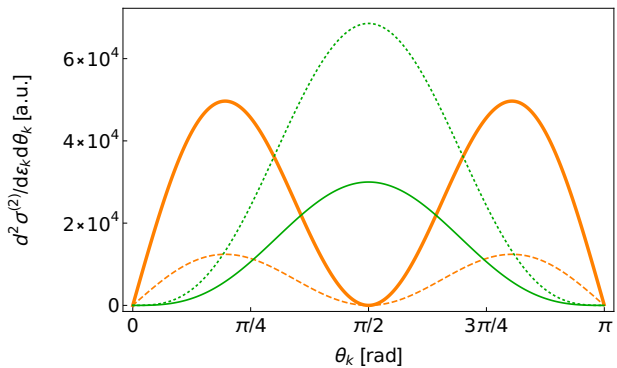}
\caption{Doubly differential cross sections $d^{2}\sigma^{(2)}/d\theta_{k} d\varepsilon_{k}|_{k=k_{0}}$ for the two-center ionization in a Li-He system, considering only the contribution from the excited state $2p_{0}$ for the orientations ${\bf R}_{\perp}$ (orange dashed) and  ${\bf R}_{\parallel}$ (orange thick), respectively. Also the contributions from the $2p_{+1}$ excited state for the same orientations are shown (green dotted, green solid). The latter curves also illustrate the case of $2p_{-1}$ since it gives the same contribution. The other parameters are as in Fig.~ \ref{fig:Fig4}.}
\label{fig:Fig5}\end{figure}
Figure \ref{fig:Fig5} shows the angular dependence of the ejected electron for different excited states of atom $B$. When only considering the excited $2p$ state with angular momentum projection $m=0$, we obtain two maxima and a deep minimum at $\theta_{k}=\pi/2$. When considering instead $m=+1$, the curves attain a single maximum at $\theta_{k}=\pi/2$. Since $m=-1$ exhibits the exactly same curve, for the interatomic orientation ${\bf R}_{\perp}$, the contribution from $m=\pm 1$ together is largely dominant and therefore resembles the corresponding angular distribution shown in Fig. \ref{fig:Fig4} which includes all excited states. Conversely, for the orientation ${\bf R}_{\parallel}$ the solid curves in Fig. \ref{fig:Fig5} show substantial contributions from all substates, which explains the appearance of the plateau shown in Fig. \ref{fig:Fig4}. Note, however, that the distributions in Fig. \ref{fig:Fig4} cannot be obtained directly from summing the respective curves in Fig. \ref{fig:Fig5}, because the former also contains interferences between the contributions from the various excited states.   
\\
\section{Conclusion}
Electron-impact induced ionization in a two-center atomic system has been studied. In this resonant 2C($e$,2$e$) process, an atom is ionized by electron-impact excitation of a neighbouring atom which subsequently deexcites via an interatomic Auger decay. It was shown to considerably affect the characteristics of the well-known direct electron-impact ionization of a single center. In particular, due to resonant 2C($e$,2$e$), the energy-differential cross section of electron-impact ionization may be enhanced tremendously in a narrow energy range. As a consequence, the two-center channel can provide a substantial contribution even  to the total cross section of electron-impact ionization. This effect can be highlighted if the yield of electrons within experimentally resolvable energy interval around the sharp resonance is collected. The two-center channel can also significantly modify the angular distribution of ejected electrons.

Our general predictions may be tested experimentally by using heteroatomic van-der-Waals molecules. Suitable candidates, for instance, could be Li-He dimers as considered here, or He-Ne dimers, as used in the recent 2CPI experiments \cite{2CPIexp}.

\section*{Acknowledgement}
This work has been funded by the Deutsche Forschungsgemeinschaft (DFG, German Research Foundation) under Grant No. 349581371 (MU 3149/4-1 and VO 1278/4-1).



\begin{thebibliography}{33}

\bibitem{book} T. D. M\"ark and G. H. Dunn (eds.), {\it Electron Impact Ionization} (Springer, Wien, 1985); A. M\"uller, Adv. At. Mol. Opt. Phys. \textbf{55}, 293 (2008).

\bibitem{Peart} B. Peart and K. T. Dolder, J. Phys. B {\bf 1}, 872 (1968).

\bibitem{Hahn} K. J. LaGattuta and Y. Hahn, Phys. Rev. A {\bf 24}, 2273(R) (1981).

\bibitem{Muller} A. M\"uller, K. Tinschert, G. Hofmann, E. Salzborn, and G. H. Dunn, Phys. Rev. Lett. {\bf 61}, 70 (1988).


\bibitem{ICD} L. S. Cederbaum, J. Zobeley, and F. Tarantelli, Phys. Rev. Lett. \textbf{79}, 4778 (1997);
V. Averbukh, I. B. M\"uller, and L. S. Cederbaum, ibid. \textbf{93}, 263002 (2004).

\bibitem{ICDres} So-called resonant ICD was studied, e.g., in K. Gokhberg, V. Averbukh, and L. S. Cederbaum, J. Chem. Phys. \textbf{124}, 144315 (2006).

\bibitem{ICDrev} For reviews on ICD, see 
R. Santra and L. S. Cederbaum, Phys. Rep. \textbf{368}, 1 (2002);
V. Averbukh {\it et al.}, J. Electron Spectrosc. Relat. Phenom. {\bf 183}, 36 (2011); 
U. Hergenhahn, {\it ibid.} {\bf 184}, 78 (2011);
T. Jahnke, J. Phys. B \textbf{48}, 082001 (2015).

\bibitem{dimers} T. Jahnke \textit{et al.}, Phys. Rev. Lett. \textbf{93}, 083002 (2004);
Y.~Morishita \textit{et al.}, ibid. \textbf{96}, 243402 (2006);
T.~Havermeier \textit{et al.}, ibid. \textbf{104}, 133401 (2010).

\bibitem{clusters} S. Marburger, O. Kugeler, U. Hergenhahn, and T. M\"oller,
Phys. Rev. Lett. \textbf{90}, 203401 (2003).

\bibitem{water} T. Jahnke \textit{et al.}, Nature Phys. \textbf{6}, 139 (2010);
M. Mucke \textit{et al.}, ibid. \textbf{6}, 143 (2010).

\bibitem{2CPI} B. Najjari, A. B. Voitkiv, and C. M\"{u}ller, Phys. Rev. Lett. {\bf 105}, 153002 
(2010); A. B. Voitkiv and B. Najjari, Phys. Rev. A \textbf{84} 013415 (2011).

\bibitem{Perina}
J. Pe\v{r}ina, A. Luk\v{s}, V. Pe\v{r}inov{\'a}, and W. Leo{\'n}ski, Phys. Rev. A {\bf 83}, 053416 (2011);
V. Pe\v{r}inov{\'a}, A. Luk\v{s}, J.~K\v{r}epelka, and J. Pe\v{r}ina, ibid. \textbf{90}, 033428 (2014).

\bibitem{2CPIexp}
F. Trinter \textit{et al.}, Phys. Rev. Lett. \textbf{111}, 233004 (2013);
A.~Mhamdi \textit{et al.}, Phys. Rev. A \textbf{97}, 053407 (2018).

\bibitem{Hergenhahn} A. Hans, P. Schmidt, C. Ozga, C. Richter, H. Otto, X. Holzapfel, G. Hartmann,
A. Ehresmann, U. Hergenhahn, and A. Knie, J. Phys. Chem. Lett. {\bf 10}, 1078 (2019)

\bibitem{Lanzhou}
S. Yan \textit{et al.}, Phys. Rev. A \textbf{88}, 042712 (2013);
S. Yan, P. Zhang, X. Ma, S. Xu, S. X. Tian, B. Li, X.~L.~Zhu, W.~T.~Feng, and D. M. Zhao,
ibid. \textbf{89}, 062707 (2014).

\bibitem{Lanzhou2}
S. Yan, P. Zhang, V. Stumpf, K. Gokhberg, X. C. Zhang, S. Xu, B. Li, L.~L.~Shen, X.~L.~Zhu, W.~T.~Feng, S. F. Zhang, D. M. Zhao, and X.~Ma, Phys. Rev. A \textbf{97}, 010701(R) (2018).

\bibitem{Dorn} T. Pfl\"uger, X. Ren and A. Dorn, Phys. Rev. A \textbf{91}, 052701 (2015);
X. Ren, E. J. Al Maalouf, A. Dorn and S. Denifl, Nature Commun. \textbf{7}, 11093 (2016).

\bibitem{Grieves} G. A. Grieves and T. M. Orlando, Phys. Rev. Lett. \textbf{107}, 016104 (2011).

\bibitem{2CDR} C. M\"{u}ller, A. B. Voitkiv, J. R. Crespo Lopez-Urrutia, and Z. Harman,
Phys. Rev. Lett. {\bf 104}, 233202 (2010);
A.~B.~Voitkiv and B. Najjari, Phys. Rev. A \textbf{82}, 052708 (2010).

\bibitem{2CRS} A. Eckey, A. Jacob, A. B. Voitkiv, and C. M\"uller, Phys. Rev. A {\bf 98}, 012710 (2018).

\bibitem{ICEC} K. Gokhberg and L. S. Cederbaum, J. Phys. B \textbf{42}, 231001(FTC) (2009);
Phys. Rev. A \textbf{82}, 052707 (2010).


\bibitem{Wachter}  See, e.g., A. Wachter, {\it Relativistic Quantum Mechanics} (Springer, Berlin, 2011); Sec. 3.

\bibitem{q-dots} ICEC has also been studied in a system of two quantum dots, see
F. M. Pont, A. Bande, and L. S. Cederbaum, Phys. Rev. B \textbf{88}, 241304(R) (2013);
J. Phys.: Cond. Matter \textbf{28}, 075301 (2016).

\bibitem{Samson}  
J.A.R Samson,{\it The Measurement of the Photoionization Cross Sections of Atomic Gases},(Space Sciences Laboratory. G C A Corporation. Bedford)


\bibitem{NIST} Atomic spectra database of the National Institute of Standards and Technology (NIST), available at https://www.nist.gov/pml/atomic-spectra-database

\bibitem{Shyn} T.W. Shyn, Phys. Rev. A {\bf 45}, 5 (1992)

\bibitem{Lotz} W. Z. Lotz, Zeitschrift fuer Physik {\bf 206}, 205-211 (1967)

\bibitem{Ganas}  
P.S. Ganas, Journal of Applied Physics {\bf 52}, 19 (1981)

\bibitem{Friedrich} B. Friedrich, Physics {\bf 6}, 42 (2013)

\bibitem{LandauQM} L. D. Landau and E. M. Lifshitz, Quantum Mechanics, (Pergamon, Oxford, 1965); see Sec.~\S\,136 and App.~\S\,e.

\bibitem{Fussnote} Contributions to resonant 2C($e$,$2e$) from excited $np$ states with $n\geq 3$ would lead to resonance peaks at correspondingly higher energies. They are not shown in Fig.~ \ref{fig:Fig2}. Their contributions to the total cross section of 2C($e$,$2e$) are suppressed. In the Li-He system at $R=20\,a_{0}$ one obtains $\sigma^{(2)}_{1s\rightarrow 3p}\approx 0.14\sigma^{(2)}_{1s\rightarrow 2p}$, for example.

\bibitem{Bhatia} A. K. Bhatia, A. Temkin, and A. Silver,
Phys. Rev. A {\bf 12}, 2044 (1975)









\end{thebibliography}
\end{document}